\documentclass[11pt.a4paper]{article}
\usepackage{jheppub}
\begin{document}
\hyphenation{phys-ics spher-oids}

\title{Predictions from extra dimensions.}

\author{Steven C. Strausz}
\affiliation{University of Washington,\\
  Box 351560, Seattle, WA 98195-1560, USA} 
\emailAdd{strausz@phys.washington.edu}

\abstract{
By representing the electroweak gauge symmetry group
${\rm SU}(2) \times {\rm U}(1)$
by a hypertorus ${\rm S}_2 \times {\rm S}_1$,
the electroweak mixing angle
and the fine structure constant are predicted.
By representing neutrinos as oscillating spheroid perturbations
on the same hypertorus,
the number of neutrino families
and the neutrino mixing matrix 
are predicted.
}

\keywords{
Kaluza-Klein Theory (Higher-dimensional Gravity),
Weinberg-Salam Electroweak Model, 
Beyond the Standard Model 
}

\maketitle

\section{Introduction}

General relativity describes the gravitational force
as being due to the curvature of space and time\cite{einstein}.
Quantum physics describes quantum forces
as being due to the exchange of intermediate fields.
How these two different descriptions of forces
are combined at the quantum level has eluded physics.
Can general relativity be applied to quantum physics
to predict experimentally measured quantities?

\section{Electroweak Mixing Angle}

Measurements of the electroweak interaction 
have revealed the massless electromagnetic photon $A_\mu$ 
and the massive neutral weak boson $Z_\mu$
to be mixtures of the neutral hypercharge field $B_\mu$, 
and the neutral third component of the weak isospin field $W_{3\mu}$\cite{glashow}\cite{weinberg}\cite{salam}
\begin{eqnarray}
\left( \begin{array}{c} A_\mu  \\ Z_\mu  \end{array} \right) 
&=&
\left( \begin{array}{rr} \cos \theta_W & \sin \theta_W  \\
                        -\sin \theta_W & \cos \theta_W  \end{array} \right) 
\left( \begin{array}{c} B_\mu  \\ W_{3\mu}  \end{array} \right) .
\end{eqnarray}
The standard model of particle physics offers no established explanation
for the value of this Weinberg electroweak mixing angle $\theta_W$.

Gravity has been explained as due to the curvature of spacetime.
The Kaluza-Klein model proposed that electromagnetism is due
to the skewing of an extra spacelike dimension curled to a small radius
with the coupling strength of electromagnetism
inversely related to the curvature radius\cite{kaluza}\cite{klein}.
The model has been extended to extra gauge interactions
by including extra dimensions with spatial symmetries matching the gauge symmetries\cite{dewitt}.
Begin with the usual spacetime dimensions $x^\mu$
described by the spacetime metric $g_{\mu\nu}(x)$. 
Add extra dimensions $y^m$
described by extra dimensions metric $h_{mn}(y)$,
Killing vectors $K_{am}(y)$,
perturbed vector fields $A_{a\mu}(x)$,
and perturbed scalar fields $S_{mn}(x)$
\begin{eqnarray}
g_{\mu\nu}(x)  &\rightarrow&
\left( \begin{array}{ll}
  g_{\mu\nu}(x)
  \,+\, (h^{mn}(y)+S^{mn}(x))\,K_{cm}(y)\,A_{c\mu}(x)\, K_{dn}(y)\,A_{d\nu}(x)\;\;
  & K_{am}(y)\, A_{a\nu}(x) \\       
    K_{bm}(y)\, A_{b\mu}(x)\;\; & h_{mn}(y)+S_{mn}(x)
\end{array} \right) .
\end{eqnarray}
Perhaps the mixing angle is specific to our universe,
and other universes in the multiverse
have different mixing angles and different interactions\cite{multi},
or perhaps the mixing angle can offer insight into physics beyond the standard model.
Can the mixing angle be predicted from physics beyond the standard model
such as theories with extra dimensions?

\section{Interactions and Geometries}

The electroweak gauge group can be described by the
gauge transformations of the product group ${\rm SU}(2) \times {\rm U}(1)$.
The one transformation of the group ${\rm U}(1)$
can be trivially described by the rotation of the
one dimensional circular space ${\rm S}_1$.
The three transformations of the group ${\rm SU}(2)$
can be described by the three rotations of
the one dimensional complex projective line ${\rm CP}_1$,
or the two dimensional real projective Riemann sphere ${\rm P}_2$.
The nonorientable Riemann sphere with constant curvature can be described
by the orientable two dimensional real sphere ${\rm S}_2$ but with its antipodes identified.
The total integration of the local curvature is relevant.
Double covering the Riemann sphere does not alter the predicted mixing angle.

Consider the closed three dimensional real space
described by coordinates $y_m=(y_1, y_2, y_3)$ 
constructed from a circle ${\rm S}_1$ with radius $r_1$
and described by an circular angle $0<y_1<2\pi$,
around which is revolved
a two dimensional surface of a sphere $S_2$ with radius $r_2$
with circle radius larger than sphere radius $r_1>r_2$
described by latitude angle $0<y_2<\pi$
and longitude angle $0<y_3<2\pi$
with the hypertorus metric
\begin{eqnarray}
h_{mn}(r_1,r_2) =
\left( \begin{array}{ccc} 
        \left( r_1 \,-\, r_2\, \sin y_2\, \cos y_3 \right)^2 & 0 & 0 \\
0 &     \left(           r_2                      \right)^2     & 0 \\
0 & 0 & \left(           r_2\, \sin y_2            \right)^2           
        \end{array} \right) 
\end{eqnarray}
with the hypertorus Jacobian integration factor
\begin{eqnarray}
\sqrt{|h|} &=& ( r_1\, -\,r_2 \;\sin y_2\; \cos y_3 )\; r_2^{\,2}\; \sin y_2 .
\end{eqnarray}
From the metric is computed the three dimensional spatial volume
\begin{eqnarray}
  H(r_1,r_2) &=& \int_{0}^{2\pi} dy_1\; \int_{0}^{\pi} dy_2\; \int_{0}^{2\pi} dy_3\; \sqrt{|h|} .
\end{eqnarray}
The traditional Einstein-Hilbert action consists of a Ricci scalar term.
The addition of a Ricci tensor squared term has been studied by others\cite{ricci}\cite{ricci2}.
From the metric are computed the usual quantities of general relativity,
Affine connection, Riemann tensor,
and the Ricci curvature tensor $R_m^{\;\;n}$
which describes the curvature of space\cite{weinberg2}.
Calculate the Ricci tensor squared
\begin{eqnarray}
R_m^{\;\;n}\; R_n^{\;\;m}  
  &=&    \frac { r_2^2\,\sin y_2\,(1\,+\,\sin^4 y_2)\, (r_1\,-2\, r_2\, \sin y_2\,\cos y_3)^2}
          { r_1\,-\, r_2\, \sin y_2\,\cos y_3 }\;+\; \nonumber\\
  & &\;+\,4\, \sin^3 y_2\, \cos^2 y_3\, (r_1\,-\, r_2\, \sin y_2\, \cos y_3)^3 .
\end{eqnarray}
Integrate the Ricci tensor squared for the Lagrangian
\begin{eqnarray}
  L(r_1, r_2) &=& \frac{1} {16\pi G}\;
  \int_{0}^{2\pi} dy_1\; \int_{0}^{\pi} dy_2\; \int_{0}^{2\pi} dy_3\;
            \sqrt{|h|}\; R_m^{\;\;n}\; R_n^{\;\;m}.                             
\end{eqnarray}
For this computation without fixed constant parameters,
only the ratio of the two radii $r_1/r_2$ matters,
not the magnitude of each radius.
To find the extremal radii ratio of the ground state of the curved space,
the extremal radii ratio is computed
by holding constant the spatial volume while minimizing the Lagrangian 
using the Lagrange multiplier method
\begin{eqnarray}
        \frac{\partial H} {\partial r_1} 
        \frac{\partial L} {\partial r_2} 
  \;-\; \frac{\partial H} {\partial r_2} 
        \frac{\partial L} {\partial r_1} 
      &=& 0.
\end{eqnarray}
The extremal radii ratio for minimum curvature is computed
\begin{eqnarray}
  r_1/ r_2 &=& 1.1808\;\;.
\end{eqnarray}
The extremal radii ratio represents the ground state of the unperturbed space.
The electroweak vector fields represent the perturbations of the ground state.
Consider the mixing angle as a measure 
of the hypercharge circle radius and the weak sphere radius dependence
on each other to perturbation.
A vanishing mixing angle would indicate full independent nonmixing
between hypercharge and weak third component.
A unit tangent mixing angle would indicate an equal sharing
between the two vector fields.
The electroweak mixing angle is predicted to be
\begin{eqnarray}
  \sin^2 \theta_{W\;{\rm theory}}
  &=& \sin^2\left( \arctan\, P \right) \nonumber \\
  &=& 0.2324\;\; .
\end{eqnarray}
The electroweak mixing angle runs with the energy scale.
The measured mixing angle is further complicated 
by renormalized quantum corrections
and uncertainties of standard model parameters
which slightly alter the angle.
At the low energy scale,
the electroweak mixing angle is experimentally measured to be\cite{low}
\begin{eqnarray}
\sin^2 \theta_{W}(m=0)_{\;{\rm experiment}} &=& 0.23867 \pm 0.00016\;\; .
\end{eqnarray}
At the energy scale of the $Z^0$ boson mass,
the electroweak mixing angle with modified minimal subtraction
is experimentally measured to be\cite{pdg}
\begin{eqnarray}
\sin^2 \theta_{W}(m=m_{Z^0})_{\;\rm experiment}  &=&  0.23122 \pm 0.00003\;\; .
\end{eqnarray}
This predicted electroweak mixing angle agrees with 
the measured electroweak mixing angle better than one percent.

\section{Neutrino Flavors}

Unlike spin and charge,
neutrino flavor carries no conserved quantum numbers.
Neutrinos $\psi_i$ interact in their flavor eigenstates
$\nu_e,\, \nu_\mu,\, \nu_\tau$
and propagate in their mass eigenstates
$\nu_1,\, \nu_2,\,  \nu_3$
allowing neutrinos to oscillate among eigenstates
described by the unitary matrix $U_{ij}$.
No generally accepted explanation exists why
the number of neutrino eigenstates equals three.
Consider the Lagrangian neutrino kinetic term, 
weak interaction term, and mass term
\begin{eqnarray}
  \cal{L} &=& \overline\psi_i\, \gamma^\mu\, \left(1-\gamma^5\right)
  \left[ i\, \partial_\mu 
        \;+\;g\,  Z^0_\mu\,U_{ij}\right]   \psi_j
        \;+\; m_i\; \overline\psi_i\; \psi_i
    \end{eqnarray} 
with spinor indices hidden.
Extend the neutrino spinor field $\psi$ beyond 4-spacetime
to include extra dimensions
\begin{eqnarray}
\psi(x^\mu) &\rightarrow& \psi(x^\mu,y^m) .
\end{eqnarray}
Perturbations $p_i(y^m)$ in the extra dimensions
can act independently of the 4-spacetime dimensions
\begin{eqnarray}
\psi_i(x^\mu,y^m) &\rightarrow& \psi(x^\mu)\;  p_i(y^m) .
\end{eqnarray}
Let three extra dimensions form a hypertorus product space
of a circle $S_1$ with coordinate $y_1$
and a sphere $S_2$ with coordinates $y_2,y_3$.
Consider neutrinos as small perturbations of spheres
oscillating between
elongated prolate spheroids and flattened oblate spheroids.
A sphere with three orthogonal rotational axes
allows three independent orthogonal oscillating spheroids.
These perturbations carry no conserved quantities such as
linear or rotational momenta
which would inhibit neutrino oscillations.
The number of theoretically predicted independent perpendicular perturbations
\begin{eqnarray}
N_{\;{\rm theory}} &=& 3
\end{eqnarray}
agrees well with
the number of the experimentally measured neutrino flavors
\begin{eqnarray}
  N_{\;{\rm experiment} } &=& 3
\end{eqnarray}
although small numbers carry little statistical significance.

\section{Oscillating Spheroid Perturbations}

Let 4-spacetime be expanded with three
extra spatial dimensional coordinates $y^m=(y^1,y^2,y^3)$,
with the same hypertorus product space metric 
\begin{eqnarray}
h_{mn} &=& \left( \begin{array}{ccc}
    (r_1 -r_2\, \sin y_2\, \cos y_3 )^2  & 0 & 0   \\
    0 &   (r_2)^2  & 0      \\    
    0 & 0 &  (r_2\, \sin y_2)^2 
    \end{array} \right)
\end{eqnarray}
with the same hypertorus product space Jacobian integration factor
\begin{eqnarray}
\sqrt{|h|} &=& ( r_1\, -\,r_2 \;\sin y_2\; \cos y_3 )\; r_2^{\,2}\; \sin y_2 . 
\end{eqnarray}
Perturb the sphere radius
\begin{eqnarray}
  r_2 &\rightarrow& r_2\;+\;p_i
\end{eqnarray}
with three small amplitude, perpendicular spheroids $p_i(y^m)$
oscillating between elongated prolate and flattened oblate.
To preserve the surface area of the perturbed oscillating spheroids,
the expansion or contraction
of the polar radius of the prolate spheroid
compared with that of a unit amount
in the oblate equatorial radius
is about $q=2$.\cite{spheroid}.
To avoid creating electric charge
from momentum around the circle,
let the three perpendicular oscillating spheroids
all travel in both directions around the circle coordinate $y^1$
with the approximations            
\begin{eqnarray}
  p_1 &=& \left[\,  q\, \cos(r_2 y_2)\,                                \cos^2 t
                  \,-\, \sin(r_2 y_2)\,                             (1-\cos^2 t)\, \right]\;\times \nonumber \\
      &&  \;\;\; \times\,\left[\,\sin(2 r_1 y_1 -t)\,+\, \sin(2 r_1 y_1 + t)\right]/2  \\
  p_2 &=& \left[\,  q\, \cos(r_2 (y_1+y_3)+\phi_2)\,  \sin(r_2 y_2)\,   \cos^2 t\, \right.\; + \nonumber \\
      &&  \;\;\; \left.  -\,  \sin(r_2 (y_1+y_3)+\phi_2)\,    (1-\cos^2 t)\, \right]\;\times  \nonumber \\
      &&  \;\;\; \times\,\left[\,\sin(2 r_1 y_1 -t+\phi_2)\,+\, \sin(2 r_1 y_1 + t+\phi_2)\, \right]/2  \\
  p_3 &=& \left[\,  q\, \sin(r_2 (y_1+y_3)+\phi_3)\,  \sin(r_2 y_2)\,   \cos^2  t\, \right.\;+ \nonumber \\
      &&  \;\;\; \left.   -\,  \cos(r_2 (y_1+y_3)+\phi_3) \,       (1-\cos^2 t)\, \right]\;\times \nonumber \\ 
      &&  \;\;\; \times\,\left[\,\sin(2 r_1 y_1 -t+\phi_3)\,+\, \sin(2 r_1 y_1 + t+\phi_3)\, \right]/2 \;\;. 
\end{eqnarray}
The perturbations have relative ring phase shifts $\phi_2, \phi_3$
to allow their maximums to be independently set which will be minimized.
The overlapping perturbations will be integrated with their time independent perturbation counterparts
\begin{eqnarray}
p'_1 &=& \left[\, q\, \cos(r_2 y_2)\, \,-\, \sin(r_2 y_2)\, \right]\; \sin(2 r_1 y_1 )                  \\
p'_2 &=& \left[\, q\, \cos(r_2 (y_1+y_3)+\phi_2)\, \sin(r_2 y_2)\, -\,\sin(r_2 (y_1+y_3)+\phi_2)\, \right]
                      \sin(2 r_1 y_1 +\phi_2) \\
p'_3 &=& \left[\, q\, \sin(r_2 (y_1+y_3)+\phi_3)\, \sin(r_2 y_2)\, -\,\cos(r_2 (y_1+y_3)+\phi_3)\, \right]
                      \sin(2 r_1 y_1 +\phi_3) .
\end{eqnarray}
Provide uniform perturbations by integrating the perturbations for
perturbation normalization factors
\begin{eqnarray}
  a_i &=&  \left(\; \int^{2\pi}_0 dt\;   \int^{2\pi}_0 dy_1\;
                    \int^{\pi}_0 dy_2 \; \int^{2\pi}_0 dy_3\;
  \sqrt{|h|}\; p_i\; p_i \;  \right)^{1/2} .
\end{eqnarray}
The variations in the perturbation normalization factors
around the region of interest are typically a few percent.
The normalization factors have an insignificant effect on the result.

\section{Neutrino Mixing Matrix}

The Pontecorvo-Maki-Nakagawa-Sakata mixing matrix,
which henceforth will be called the neutrino mixing matrix,
describes the transition between neutrino eigenstates.
No generally accepted explanation exists why
the elements of the neutrino mixing matrix
have the values they hold.\cite{mix}
Complicating the marriage
between general relativity and quantum physics
is that general relativity lives in real numbers,
while quantum physics enjoys complex numbers.
Although the neutrino mixing matrix is a complex-valued unitary matrix,
a real-valued orthogonal matrix will be calculated.
Perhaps the complexity is a small effect
due to another physical mechanism.
Can perturbations of oscillating spheroids explain
the neutrino mixing matrix?

Imagine the set of three oscillating spheroids being
the three mass eigenstates.
Differentiate the extra dimension perturbations 
\begin{eqnarray}
  \partial p_i &=&  -\,\partial_t p_i     \,+\,\partial_{y_1} p_i
                  \,+\,\partial_{y_2} p_i  \,+\,\partial_{y_3} p_i \\
  \partial p'_i &=&  -\,\partial_t p'_i     \,+\,\partial_{y_1} p'_i
                   \,+\,\partial_{y_2} p'_i  \,+\,\partial_{y_3} p'_i .
\end{eqnarray}
Substitute the perturbations and discard higher order terms
$p'_i\,\partial p_j\rightarrow 0$
\begin{eqnarray}
  && \left( r_2 \;+\;p'_i\right)\left( r_2 \;+\;\partial p_j\right)
  \, -\,r_2^2\; +\;{\rm other\;\, combinations} \nonumber \\
&&\;\;\;\;\; \approx \left( p_i \,+\,p_j \,+\,p'_i\, +\,p'_j
  \,+\,\partial p_i\,+\,\partial p_j\,
  +\,\partial p'_i\,+\,\partial p'_j \right) r_2 .
\end{eqnarray}
The sphere radius $r_2$ is a constant.
Integrate the overlapping perturbations kinetic matrix
\begin{eqnarray}
  K_{ij} &=&  \frac{1}{( a_i a_j+a_j a_i)/2}\;
   \int^{2\pi}_0 dt \; \int^{2\pi}_0 dy_1\;  \int^{\pi}_0 dy_2\;  \int^{2\pi}_0 dy_3 
  \; \sqrt{|h|}\;\times                               \nonumber \\
&&\;\;\; \times
  \left(\, p_i \,+\,p_j \, +\,p'_i \,+\,p'_j
            \,+\,\partial p_i  \,+\,\partial p_j
            \,+\,\partial p'_i \,+\,\partial p'_j \right) r_2 .
\end{eqnarray}
Find the relative ring phase shifts $\phi_2, \phi_3$ 
that minimize the overlapping kinetic matrix squared $K_{ij} K^{ji}$.
A minimum of relative ring phase shifts is found near
\begin{eqnarray}
  \phi_2 = -0.43 \pi, \;\;\;\;\; 
  \phi_3 = +0.53 \pi .
\end{eqnarray}
Compute the phase-shifted overlap of the oscillating spheroids.
Normalize the overlapping kinetic matrix with its first element $|K_{11}|=1$
\begin{eqnarray}
 \frac{K_{ij}} {|K_{11}|}  &=&
 \left( \begin{array}{rrr}
 -1.000 & -0.403 & -0.436  \\   
 -0.403 & -0.573 &  0.176  \\ 
 -0.436 &  0.176 &  0.143 
\end{array}\right) .
\end{eqnarray}
Although this matrix is symmetric,
it is not unitary which a good neutrino mixing matrix needs to be.
Use QR decomposition with the Gram Schmidt method
to transform the kinetic matrix $K_{ij}$ through a series of approximations
into an orthogonal neutrino mixing matrix\cite{qr}\cite{qr2}
\begin{eqnarray}
Q_R\left( \frac{K_{ij}}{ |K_{11}|} \right)  &=&
\left( \begin{array}{rrr}
  -0.86 & -0.35 & -0.37  \\   
  0.02 & -0.75 &  0.66  \\ 
 -0.51 &  0.56 &  0.65 
\end{array}\right) .
\end{eqnarray}
Take absolute values which only are measured by experiment.
Since the original matrix was symmetric,
its QR decomposition can be transposed.
Exchange the second and third columns and rows,
which were arbitrarily assigned to the three perturbations
\begin{eqnarray}
\left| U_{ij}\right|_{\;{\rm theory}}  &=& 
\left( \begin{array}{rrr}
 0.86 &  0.51 &  0.02  \\   
 0.37 &  0.65 &  0.66  \\ 
 0.35 &  0.56 &  0.75 
\end{array}\right) .
\end{eqnarray}
Notice the diagonal preference of the matrix.
Many experiments have contributed measurements to the neutrino mixing matrix.
Compare with the {\sc Nu-fit 2.3} values for the $3\sigma$ confidence limits
on the experimentally measured absolute values
of the neutrino mixing matrix\cite{nufit}
\begin{eqnarray}
\left|  U_{ij} \right|_{\;{\rm experiment}}  
=
\left( \begin{array}{rrr}
      0.799   \leftrightarrow 0.844\;\;
  &   0.516   \leftrightarrow 0.582\;\;
  &   0.141   \leftrightarrow 0.156  \\
      0.242   \leftrightarrow 0.494\;\;
  &   0.467   \leftrightarrow 0.678\;\;
  &   0.639   \leftrightarrow 0.774  \\ 
      0.284   \leftrightarrow 0.521\;\;
  &   0.490   \leftrightarrow 0.695\;\;
  &   0.615   \leftrightarrow 0.754
\end{array} \right) .
\end{eqnarray}
Most of the elements of
the theoretically predicted neutrino mixing matrix 
appear within these confidence limits
of the experimentally measured neutrino mixing matrix.
The small upper right element $U_{13}$ is usually
associated with CP violating effects.

\section{Fine Structure Constant}

Consider the fine structure constant
\begin{eqnarray}
 \alpha &=& \frac{e^2}{4\,\pi\, \varepsilon_0\,  \hbar\, c}.
\end{eqnarray}
In Kaluza-Klein theory,
electric charge is the momentum
in the extra dimension quantized in the circumference,
hence electric charge
is related to the inverse of the curvature of the extra dimension. 
Compare the squared inverse curvature of the hypertorus ring
with that of a hypersphere of same dimension.
Use the same radii ratio $r_1/r_2 =1.1808$
of ring radius to sphere radius,
and the unit sphere radius $r_2=1$.
The three dimensional hypertorus Ricci scalar curvature $R_T$ is
\begin{eqnarray}
R_T &=& {
    \frac  {\left(1+\sin^2 y_2\right)  \left( r_1 + 2\, r_2\,\sin y_2\,\cos y_3  \right)}
                               { r_1 + r_2\,\sin y_2\,\cos y_{3}     }
   +\frac {           2\,\sin y_2\,\cos y_3\,
     \left(r_1 + r_2\,\sin y_2\,\cos y_3  \right) }
           {r_2} }
\end{eqnarray}
with the same hypertorus Jacobian integration factor $\sqrt{|h|}$. 
Integrate the hypertorus Ricci scalar curvature $R_T$
\begin{eqnarray}
 T &=& \int^{2\pi}_0 dy_1\; \int^{\pi}_0 dy_2\; \int^{2\pi}_0 dy_3 
       \; \sqrt{|h|}\; R_T                        \nonumber \\
   &=& 279.7\; r_2^3.
\end{eqnarray}
Compare with the three dimensions of the
hypersphere which has the Ricci scalar curvature $R_S$
\begin{eqnarray}
 R_S &=& 2\left( \sin^2 y_1\, \sin^2 y_2\, +\, \sin^2 y_1\, +\,1 \right)
\end{eqnarray}
and the hypersphere Jacobian integration factor
\begin{eqnarray}
\sqrt{|h_S|} &=& r_2^3\, \sin^2 y_1\, \sin y_2 .
\end{eqnarray}
Integrate the hypersphere Ricci scalar curvature $R_S$
\begin{eqnarray}
 S &=& \int^{\pi}_0 dy_1\; \int^{\pi}_0 dy_2\; \int^{2\pi}_0 dy_3 
       \; \sqrt{|h_S|}\; R_S                          \nonumber \\
   &=& 88.83\; r_2^3.
\end{eqnarray}
Note the difference of angles and integration limit.
With vacuum electric permittivity $\varepsilon_0=1$
and this calculation being relativistically quantum $\hbar c=1$,
calculate the fine structure constant
\begin{eqnarray}
  \alpha_{\;{\rm theory}}
    &=&  \frac{e^2}{4\,\pi\, \varepsilon_0\,  \hbar\, c} 
   \;=\; \frac{1}{4\,\pi} \left( \frac{S}{T}\right)^2
                              \nonumber \\
    &=&  \frac{1}{124.6}\;\;.
\end{eqnarray}
The fine structure constant runs with the energy scale.
At the low energy scale,
the fine structure constant is experimentally measured to be\cite{pdge}
\begin{eqnarray}
  \alpha(m=0)_{\;{\rm experiment}}
  &=& \frac{1}{137.035 999 139 \pm 0.000 000 031}\;\;.
\end{eqnarray}
At the momentum transfer energy scale of the $Z^0$ boson mass,
the effective fine structure constant
is experimentally measured to be\cite{alphamz}
\begin{eqnarray}
  \alpha(m=m_{Z^0})_{\; {\rm experiment}}
  &=& \frac{1}{128.936 \pm 0.046}\;\;.
\end{eqnarray}
Theory agrees with experiment at the $Z^0$ boson mass scale
within a few percent
although disagrees enormously 
with the precise experimental error confidence limits.

\section{Conclusions and Discussions}

The electroweak mixing angle was predicted within
an accuracy of about $1$ percent.
The number of neutrino flavors was predicted precisely
although with small number statistics.
Most of the elements of the neutrino mixing matrix were predicted
within $3\sigma$ of their experimental measurements.
The fine structure constant was predicted within a few percent.
Decreasing the statistical significance of these predictions
is the number of trials performed with different assumptions.
Perhaps general relativity with extra dimensions
can offer predictive value for quantum physics.


\end{document}